\documentclass[aps,pra,twocolumn,superscriptaddress]{revtex4-2}

\usepackage{graphicx}
\usepackage{siunitx}
\usepackage[hyperfootnotes=false,linkbordercolor={0 .298 .6}, citebordercolor={.8 0 0}, urlbordercolor={0 .298 .6}]{hyperref}
\usepackage{csquotes}

\hypersetup{
	colorlinks,
	linkcolor=[rgb]{0,.298,.6},
	citecolor=[rgb]{0,.298,.6},
	urlcolor=[rgb]{0,.298,.6}
}

\begin{document}

\title{One- and \color{black} two-photon spectroscopy  with \color{black} a  test \color{black} of the Kennard-Stepanov relation in high-pressure two-species xenon-noble gas mixtures}

\author{Eric Boltersdorf}\email[]{boltersdorf@iap.uni-bonn.de}

\author{Thilo vom H{\"o}vel}
\author{Jeremy Andrew Morín Nenoff}
\author{Frank Vewinger}
\author{Martin Weitz}
\affiliation{Institut f{\"u}r Angewandte Physik, Universit{\"a}t Bonn, Wegelerstra{\ss}e 8, 53115 Bonn, Germany}

\date{\today}

\begin{abstract}
	
	Between the absorption and the emission spectral lineshapes of dense atomic and molecular media, such as dye solutions and alkali-noble buffer gas mixtures at high pressure, in many cases there exists a universal scaling, the Kennard-Stepanov relation, which is a manifestation of detailed balance. This relation plays a crucial role in recent Bose-Einstein condensation experiments of visible-spectral-photons in e.g.~dye-solution-filled optical microcavities. It has recently been proposed to use high-pressure xenon-noble gas mixtures as a thermalization medium for vacuum-ultraviolet regime photons, so as to extend the achievable wavelength range of such Bose-Einstein-condensed optical sources from the visible to the vacuum-ultraviolet regime. In this work, we report two-photon excitation spectroscopy measurements of ground state ($5p^6$) xenon atoms subject to up to \SI{80}{\bar} of helium or krypton buffer gas pressure, respectively, in the \SI{220}{\nano\meter} - \SI{260}{\nano\meter} wavelength range. The study of such two-photon spectra is of interest e.g.~for the exploration of possible pumping schemes of a future vacuum-ultraviolet photon Bose-Einstein condensate. We have also recorded absorption and emission spectra of the $5p^6 \leftrightarrow 5p^56s$ single-photon transition near \SI{147}{\nano\meter}  wavelength of xenon atoms subject to \SI{80}{\bar} of krypton buffer gas pressure. We find that the ratio of absorption and emission  shows \color{black} a Kennard-Stepanov scaling, which suggests that such gas mixtures are promising candidates as a thermalization medium for a Bose-Einstein condensate of vacuum-ultraviolet photons.
\end{abstract}
\maketitle

\section{Introduction}\label{sec: introduction}
	Optical spectroscopy of gases subject to high pressure has long been of interest in plasma physics and the astrophysics of stellar atmospheres \cite{Khalafinejad2017, Allard2006,Leggett2012, Uzdensky2014,Martin1960}. While in usual atomic and molecular gas phase spectroscopy conditions are such that excitation causes far-from-thermal-equilibrium conditions, at sufficient gas density, provided that interparticle collisions are elastic, lineshapes can be governed by thermodynamic scaling laws \cite{KennardFluorescence1918, StepanovFluorescence1957, KashaCharacterization1950}. In particular, between the absorption $\alpha(\omega)$ and the emission $f(\omega)$ lineshapes of absorbers a Boltzmann-type frequency scaling of the form $\frac{f(\omega)}{\alpha(\omega)} \propto \omega^3\exp{\left(-\frac{\hbar\omega}{k_{\textrm{B}}T}\right)}$, the so-called Kennard-Stepanov relation, \color{black} has been observed, where $T$ is the sample temperature and $\omega$ the optical frequency \cite{MoroshkinKennardStepanov2014}. This relation, which leads to a Stokes-shifted emission, \color{black} is understood to stem from detailed balance in a system with a substructure (for gaseous systems the quasimolecular collisional manifolds) in the electronic ground and excited states respectively being in thermal equilibrium \cite{Sawicki1996}. The Kennard-Stepanov relation is long known to be applicable for some dye molecules in liquid solution \cite{StepanovFluorescence1957}, and also holds for some semiconductor systems \cite{RoosbroeckShockley1954}, where it is more commonly known as the van Roosbroeck-Shockley relation. It plays a key role in recent experiments on the Bose-Einstein condensation of photons in material-filled optical microcavities, which at present operate in the visible or near-infrared spectral regime \cite{KlaersCondensation2010, KlaersThermalization2010, MarelicCondensation2015, GrevelingDensity2018, SchmittCoherence2016}. In these experiments, the microcavity with mirror spacing in the wavelength regime provides a suitable density of states for Bose-Einstein condensation along with a low-frequency cutoff, and photons thermalize to ambient temperature by repeated absorption and reemission processes on e.g.~dye molecules.
	\begin{figure}[t]
		\includegraphics[]{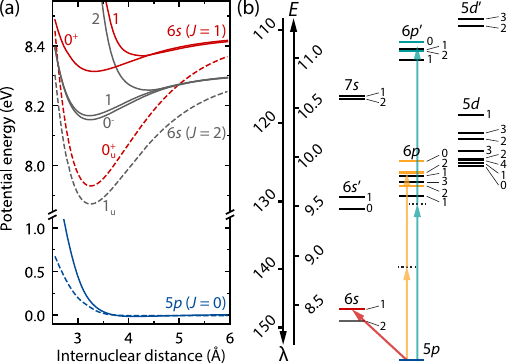}
		\caption{(a) Molecular potentials of the three lowest energetic electronic states of the xenon-krypton excimer, according to \cite{Gerasimov2004}. The dashed lines represent the corresponding potential curves for the binary xenon system. The asymptotic atomic xenon states are indicated on the right hand side, with the label $5p^5$ omitted in all states for brevity. (b) Energy level scheme of atomic xenon \cite{NISTenergyLevels2023}, with for this work relevant one- and two-photon transitions indicated. The corresponding single-photon transition wavelength is indicated on the y-axis. \color{black} The term indicates the configuration of the outermost electron and the number the corresponding total angular momentum $J$. The $5p^56s$ ($J$ = 1) state, whose fluorescence is monitored in the described experiments, is populated either by direct excitation or by collisional deactivation from higher energetic states. \label{fig:energyLevels}}
	\end{figure}
	A photon Bose-Einstein condensate constitutes a coherent light source without the necessity of inversion, where spontaneous emission is retrapped and thus can be \enquote{recycled}. This scheme makes it an attractive candidate for a coherent light source at short wavelengths, e.g.~in the vacuum-ultraviolet (VUV: $\lambda = \SI{100}{\nano\meter} \textrm{ to } \SI{200}{\nm}$) \cite{ZaidelVUV1970}, where laser operation is typically hard to achieve because of the $1/\omega^3$ scaling of excited-state lifetimes. For the thermalization of a vacuum-ultraviolet photon gas, dye molecules, as used for visible photon Bose-Einstein condensation, cannot be used due to their lack of suitable transitions. Earlier work \cite{Hoevel2023} of our group has proposed the usage of high-pressure noble gas mixtures containing xenon as the optically active constituent and a lighter noble gas, e.g.~helium or krypton, as a buffer gas. Xenon has a closed electronic transition from the $5p^6$ ground state to the $5p^56s$ excited state near \SI{146.96}{\nm} wavelength, which is the desired transition involved in a xenon-gas-based photon condensate. One possible means to pump a future vacuum-ultraviolet photon Bose-Einstein condensate is two-photon pumping, noting that powerful continuous-wave \color{black} frequency-converted laser sources are commercially available down to near \SI{200}{\nano\meter} wavelength, with this wavelength range being determined by the transparency range of nonlinear crystals.\\
	Here, we report on experimental results observing two-photon spectra of gas mixtures of xenon atoms with the lighter buffer gases krypton and helium, respectively, at buffer gas pressures of up to \SI{80}{\bar} in the \SI{220}{\nano\meter} to \SI{260}{\nano\meter} wavelength range. Besides the allowed two-photon resonances from the xenon $5p^6$ ground state to $5p^56p$ and $5p^56p'$, which are clearly observed for the case of helium buffer gas as well as for lower krypton gas pressures of a few bars, we observe further resonances which we attribute to \enquote{parity-forbidden} transitions to the $5p^57s$ and $5p^55d$ states in the dense krypton system, while the \enquote{allowed} two-photon resonances tend to vanish here. In further measurements with a xenon-krypton buffer gas mixture of \SI{80}{\bar} pressure, we have recorded absorption and emission spectra of the $5p^6 \leftrightarrow 5p^56s$ single-photon resonance near \SI{147}{\nano\meter} wavelength. Our corresponding data gives evidence for a Kennard-Stepanov scaling of the ratio between absorption and emission for this vacuum-ultraviolet-spectral-regime transition with a good spectral overlap.\\
	The solid lines in Fig.~\ref{fig:energyLevels} (a)  show \color{black} the variation of the ground and the lowest electronically excited levels of the xenon-krypton system, assuming a binary collisional model. While in the visible-spectral-range liquid dye system collisions of solvent molecules cause thermalization of the rovibrational manifold in the upper and lower electronic states, in the described gaseous systems frequent collisions among the gas atoms are expected to cause a thermalization of upper and lower electronic state quasimolecular manifolds when the collisional rate exceeds spontaneous (and inelastic) decay. Typical experimental parameters, see also earlier work mostly in the visible spectral range \cite{OckenfelsSpectroscopy2022, MoroshkinKennardStepanov2014}, are gas pressures in the range of \SI{100}{\bar}. The dashed lines in Fig.~\ref{fig:energyLevels} (a) give the quasimolecular energy levels of the binary xenon system, for which the upper electronic state (asymptotically the atomic $5p^56s$ state) is much more strongly bound as compared to the heteronuclear system, which results in strongly Stokes-shifted emission centered around \SI{172}{\nano\meter} wavelength, the so-called second excimer continuum. The large Stokes Shift relative to the \SI{147}{\nano\meter} absorption line prevents effective reabsorption of the emission in a pure xenon gas. Therefore the use of noble gas mixtures of an optically active xenon atom with a lighter gas atom, such as a xenon-krypton mixture (Fig.~\ref{fig:energyLevels} (a)), seems advantageous for use as thermalization medium in a vacuum-ultraviolet photon Bose-Einstein condensation experiment.\\
	Fig.~\ref{fig:energyLevels}(b) shows a level scheme of the atomic xenon system, with the lowest energetic (dipole-) allowed one- and two-photon transitions indicated. Upon e.g.~performing two-photon excitation of the $5p^6 \rightarrow 5p^56p$ transition with radiation near \SI{250}{\nano\meter}, upon subsequent spontaneous decay or collisional deactivation (the latter channel is expected to dominate at the here-discussed high gas pressures), the $5p^56s$ state is populated, which is the upper electronic state of the \SI{147}{\nano\meter} wavelength one-photon transition.\\	
	Available literature on spectra of heteronuclear xenon-noble gas mixtures, particularly at higher pressures, is rare. Previous measurements have determined single-photon spectra in pure xenon gas at pressures of up to \SI{160}{\bar} \cite{WahlAbsorption2018, WahlXenon2021, BorovichAbsorption1973}. Moreover, two-photon excitation measurements have been performed in pure xenon up to a pressure of \SI{95}{\bar} \cite{Hoevel2023,GornikTwoPhoton1981, RaymondTwoPhoton1984, BoweringCollisionalLifetimes1986, BoweringCollisionalState1986, WhiteheadDeactivation1995}. For the heteronuclear systems, emission spectra have been recorded by bombardment of cryogenic krypton \cite{NowakHeteronuclear1985} which gives evidence to the existence of the heteronuclear quasi-molecule, additionally in \cite{Gerasimov2000} these spectra have been shown also for an excitation by discharge. Further, the energy transfer between different excitations in various noble gas mixtures in the VUV has been investigated \cite{EfthimiopoulosExcimer1997}. Moreover, in a regime of up to \SI{0.6}{\bar} gas pressure, single-photon absorption spectra of gas mixtures of xenon with the lighter noble gases have been reported experimentally \cite{Freeman1977}.\\
	In the following, in  Sec. \ref{sec: experimentalEnvironment} the used experimental setup is described. Section \ref{sec: ExcitationSpectroscopy} gives two-photon excitation spectroscopy results for xenon-helium and xenon-krypton gas mixtures. Subsequently, in Sec. \ref{sec: KennardStepanov} spectroscopic results regarding the fulfillment of the Kennard-Stepanov relation in a xenon-krypton gas mixture are described. Finally, section \ref{sec: conclusions} gives conclusions and an outlook.

\section{Experimental Environment}\label{sec: experimentalEnvironment}
	\begin{figure}
	\includegraphics[]{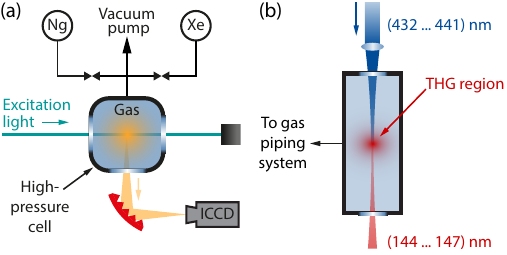}%
	\caption{(a) Schematic of the setup used for excitation spectroscopy. An exciting optical beam is sent into a high-pressure cell filled with a xenon-noble gas mixture. The resulting vacuum-ultraviolet spectral regime fluorescent radiation is guided to a grating spectrometer. For the two-photon spectroscopy work, tunable excitation light in the wavelength range \SI{220}{\nano\meter} - \SI{260}{\nano\meter} is required, as generated by frequency-doubling of the emission of a pulsed OPO laser source. (b) For the reported one-photon excitation spectroscopy work, tunable VUV radiation in the 144 to \SI{147}{\nano\meter} wavelength range is required, as is generated by frequency-tripling in a xenon-filled gas cell.
		\label{fig: experimentalEnvironment}}
	\end{figure}
	The experimental setup that we use for our excitation spectroscopy measurements is shown in Fig.~\ref{fig: experimentalEnvironment} (a) and is similar as described in earlier work of our group \cite{Hoevel2023}. The main element is a high-pressure gas cell made out of stainless steel, which provides optical access via $\textrm{MgF}_2$ windows that are transparent for VUV radiation. The cell is placed inside a vacuum chamber, which is evacuated to a pressure of around \SI{5e-5}{\milli\bar} to suppress absorption of VUV radiation by oxygen molecules in the air \cite{ZaidelVUV1970}. To determine excitation spectra, a tunable UV light source is required. Starting point for the generation of excitation light is a Q-switched pulsed laser (Spectra Physics, model Quanta Ray PRO 250-10) operating near \SI{1064}{\nm} wavelength with a repetition rate of \SI{11}{\hertz}, a pulse length of \SI{10}{\nano\second} and a pulse energy of \SI{1400}{\milli\joule}. The light is guided through a second-harmonic stage and a subsequent sum-frequency generation stage, where it is converted to pulsed radiation near \SI{355}{\nano\meter} with about \SI{400}{\milli\joule} pulse energy. This light is used to pump an optical parametric oscillator (OPO; GWU, model primoScan ULD400). It gives a tunable output between \SI{405}{\nm} and \SI{705}{\nm} wavelength. The corresponding radiation is frequency-doubled to achieve the UV radiation needed for the two-photon excitation measurements described in this work. Depending on the wavelength, the obtained UV pulse energy varies between \SI{1}{\milli\joule} and \SI{5}{\milli\joule}. Inside the vacuum chamber, the UV excitation beam is focused into the high-pressure cell by a lens of focal length \SI{25}{\milli\meter}. The emission is then collected orthogonally to the excitation beam and guided onto the entrance slit of a VUV spectrometer (H + P spectroscopy, model easyLight). Finally, the radiation is collected with an intensified CCD camera (Andor, model iStar 320T).\\
	For the part of the work investigating the Kennard-Stepanov relation of the high-pressure gas mixture (Section \ref{sec: KennardStepanov}), both single-photon absorption and emission spectra of the $5p^6 \leftrightarrow 5p^56s$ transition are recorded. For the corresponding emission measurements, the used setup is quite similar as described above, though the excitation light, which here is in the VUV spectral regime, is created by frequency-tripling of the output of the OPO laser source, which is here set to an emission between \SI{432}{\nm} and \SI{441}{\nm}. Via third harmonic generation (THG) in a gas cell, see Fig.~\ref{fig: experimentalEnvironment}(b), filled with dilute xenon gas of a few \SI{10}{\milli\bar} pressure \cite{Gray2021} we can create VUV light between \SI{144}{\nano\meter} and \SI{147}{\nano\meter} wavelength. This is subsequently used to excite the $5p^56s$ ($J = 1$) electronic state. The resulting emission is then collected as described for the case of two-photon excitation. In order to record absorption spectra, the high-pressure cell filled with the noble gas mixture is irradiated with broadband VUV light generated by a laser-induced plasma and the transmitted radiation is spectrally resolved. The corresponding setup is described in earlier work \cite{WahlXenon2021}.\\
	For all measurements described in this work, the high-pressure cell is equipped with piping systems that allow for the preparation of heteronuclear mixtures. The used gases have been delivered by the manufacturer Air Liquide (xenon: grade 4.8, helium: grade 6.0, krypton: grade 4.8).
		 
\section{Two-Photon Excitation Spectroscopy Measurements}\label{sec: ExcitationSpectroscopy}

\begin{figure}[]
		\includegraphics[]{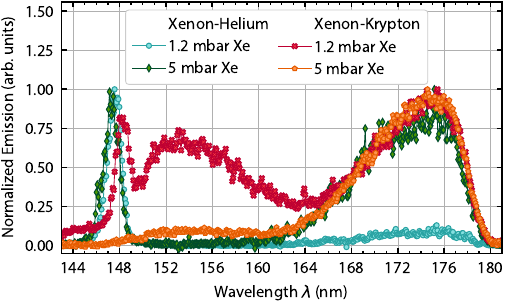}
		\caption{Exemplary xenon-helium and xenon-krypton emission spectra for a buffer gas pressure of \SI{80}{\bar} and different xenon partial pressures (see description), recorded for a two-photon excitation wavelength of \SI{248.70}{\nano\meter} in xenon-helium and \SI{231.00}{\nano\meter} in xenon-krypton  (see Fig.~\ref{fig: excitationSpectrum} for corresponding resonances in the excitation spectra) \color{black} respectively.  The data has been individually normalized to unity. \color{black} Both emission near the \SI{147}{\nano\meter} one-photon atomic xenon transition line and the xenon second excimer continuum centered at \SI{172}{\nano\meter} wavelength is visible. In contrast, for the case of the xenon-krypton mixtures, additionally emission centered near \SI{153}{\nano\meter} wavelength is visible, as this is attributed to stem from xenon- krypton excimers.\label{fig: rawSpectra}}
	\end{figure}

	In this chapter, two-photon excitation spectroscopy measurements of the $5p^6 \rightarrow 5p^56p$ and $\rightarrow 5p^56p'$ transitions, as driven with two photons of equal wavelength, on xenon-noble gas mixtures are described. Following corresponding excitation, relaxation to the $5p^56s$ ($J = 1$) state can in principle either happen by emission of an  intermediate \color{black} photon or via collisional deactivation. Rates for collisional relaxation, as have been reported for the xenon-xenon system, are at least \SI{6e-12}{\cm\cubed\per\second} at room temperature \cite{BruceRates1990}, which translates into a rate of \SI{0.14}{\per\bar\per\nano\second}. If we assume that the rate constant is not drastically different for the heteronuclear gas mixtures used here, at \SI{30}{\bar} pressure the corresponding decay is more than 100 times quicker then the rate of spontaneous emission, which is around \SI{0.033}{\per\nano\second}. Hence, we expect the transfer to the $5p^56s$ ($J = 1$) state to happen almost purely via collisional deactivation.\\
	Being in the $5p^56s$ ($J = 1$) state, a xenon atom can either undergo direct spontaneous decay, form a heteronuclear molecule with the buffer gas (helium or krypton), or form a molecule with a further, non-excited, xenon atom. In the latter case, an excited excimer xenon dimer is formed (see also the corresponding potential curves shown in Fig.~\ref{fig:energyLevels} (a)), which will spontaneously decay with the emission centered at \SI{172}{\nano\meter} wavelength. In the case of a heteronuclear molecule, the emission is expected to be closer to the atomic resonance because of a smaller potential minimum between the excited xenon and the buffer gas atom. For xenon-krypton, corresponding excimer emission has already been observed around \SI{156}{\nano\meter} in earlier work at conditions of lower pressure \cite{NowakHeteronuclear1985, Gerasimov2000}.\\
	Figure \ref{fig: rawSpectra} shows experimentally recorded spectra for mixtures of xenon-helium and xenon-krypton for a buffer gas pressure of \SI{80}{\bar} and different pressures of the optically active xenon gas respectively. For the xenon-krypton mixture at \SI{1.2}{\milli\bar} xenon gas pressure (connected red crosses), three different spectral features are visible, a spectrally relatively sharp feature near \SI{148}{\nano\meter}, which we attribute to (pressure-broadened and -shifted) atomic emission, and two broader features centered around \SI{153}{\nano\meter} and \SI{172}{\nano\meter} respectively, which are understood to stem from the emission of xenon-krypton and xenon-xenon exciplexes respectively (see also Fig.~\ref{fig:energyLevels} (a)). The data shown by the connected orange pentagons in Fig.~\ref{fig: rawSpectra} was recorded at the higher xenon contribution of \SI{5}{\milli\bar} partial pressure, for which the feature associated with the shorter-wavelength atomic emission apparently has vanished, moreover the relative contribution of the stronger Stokes-shifted peak at \SI{172}{\nano\meter} has increased. This is attributed to the larger concentration of xenon atoms, making it more probable that the energetically favored xenon excimers are formed. For the case of pure xenon atoms, the apparent reduction of the emission near the atomic resonance position at the benefit of the second excimer continuum for increased xenon concentration is well studied \cite{Ledru2007}.\\
	The connected blue circles and green diamonds give data recorded for helium buffer gas pressure at xenon concentrations of \SI{1.2}{\milli\bar} and \SI{5}{\milli\bar} partial pressures, respectively. For the corresponding data with the lighter atom buffer gas, two spectral features are visible, a peak near \SI{147}{\nano\meter} wavelength and a broad feature centered at the second excimer emission wavelength of \SI{172}{\nano\meter}. The former emission line is spectrally shifted by about \SI{1}{\nano\meter} to shorter wavelengths  with respect to the xenon-krypton data, as attributed to a different shape of the internuclear potential at large distances, near the asymptotic limit. Most notably, for the xenon-helium data only two spectral features are visible. We attribute this to the fact that the spectral emission of xenon-helium excimers, for which the molecular potentials in the excited state  due to the small mass of the helium atom \color{black} are expected to be much weaker bound than for the case of the xenon-krypton, cannot be spectrally distinguished from the atomic emission  within the experimental resolution. We have not been able to find molecular potentials of the xenon-helium system in literature, but our experimental data gives evidence for the (expected) absence of a similarly-pronounced potential minimum in the excited state as the one of the xenon-krypton system (see Fig.~\ref{fig:energyLevels} (b) for the latter). \color{black} At the higher xenon partial pressure value of \SI{5}{\milli\bar}, the relative lineshape contribution of the \SI{172}{\nano\meter} second excimer continuum stemming from the formation of xenon-xenon excimers as expected clearly increases.\\
		\begin{figure*}
		\includegraphics[]{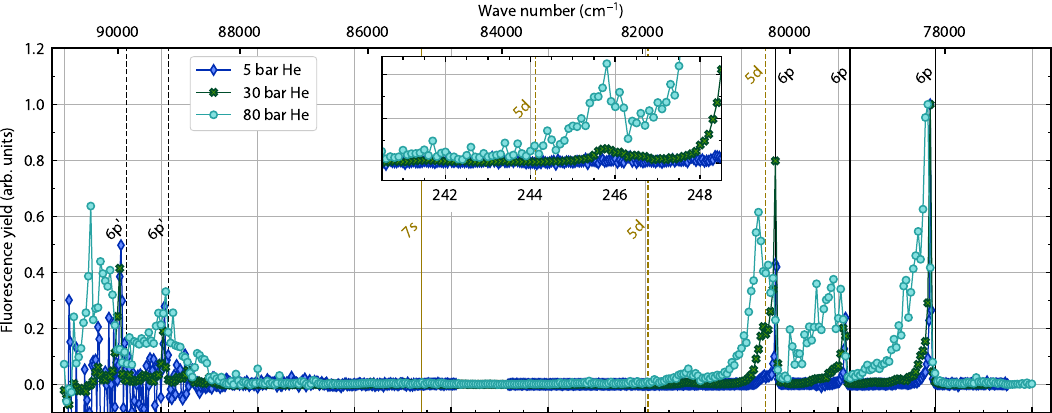}
		\includegraphics[]{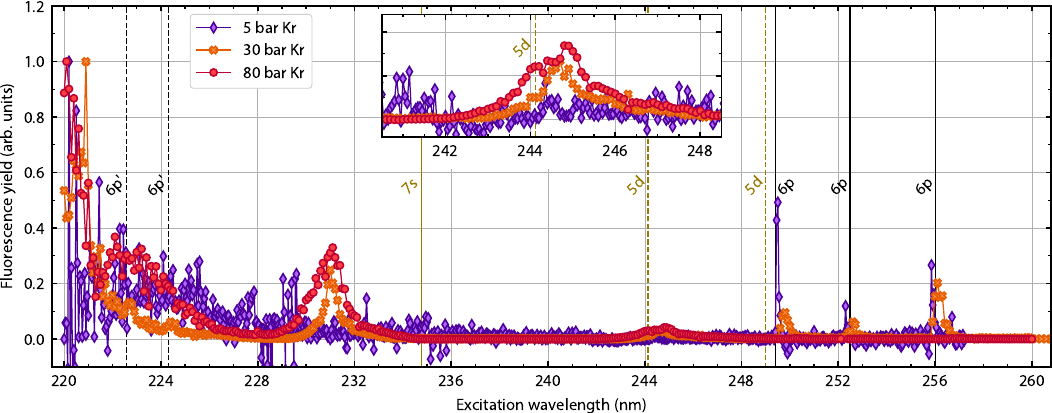}
		\caption{Two-photon excitation spectra of xenon atoms subject to helium (upper panel) and krypton (lower panel) buffer gas. The used xenon partial pressure was \SI{5}{\milli\bar}. Spectra are shown for \SI{5}{\bar}, \SI{30}{\bar} and \SI{80}{\bar} pressure of the respective buffer gas. The expected position of the unperturbed xenon atomic transitions are indicated, with $5p \rightarrow 6p$ and $6p'$ upper state configuration corresponding to allowed two-photon transitions and both $5p \rightarrow 5d$ and $7s'$ to parity-forbidden two-photon transitions. The insets give experimental data in the \SI{240.5}{\nano\meter} to \SI{248.5}{\nano\meter} range on a magnified scale, where for the case of the xenon-krypton data the observed resonance near \SI{244.5}{\nano\meter} for the \SI{30}{\bar} and \SI{80}{\bar} data give evidence for such \enquote{parity-forbidden} transitions. \label{fig: excitationSpectrum}}
	\end{figure*}To take two-photon excitation spectra, emission spectra as shown in Fig.~\ref{fig: rawSpectra} are recorded for different excitation laser wavelengths in the range between \SI{220}{\nano\meter} and \SI{260}{\nano\meter} in steps of \SI{0.1}{\nano\meter}, and the respective total fluorescence yield is determined by integration of the obtained emission spectrum for each of the applied excitation laser beam wavelengths. Further, as the output pulse energy of our used OPO varies with wavelength, the obtained data for the fluorescence yield is scaled by the respective quadratic pulse energy to match the quadratic power scaling of two-photon excitation \cite{GoeppertMayerQuadratic1931}, see also earlier work of our group for a more detailed description and characterization of this procedure \cite{Hoevel2023}.\\
	The top panel of Fig.~\ref{fig: excitationSpectrum} shows obtained two-photon excitation spectra for different values of helium buffer gas pressure and a xenon pressure of \SI{5}{\milli\bar}. In the xenon-helium data, the observed resonances near \SI{256}{\nano\meter}, \SI{252}{\nano\meter}, \SI{249}{\nano\meter}, \SI{224.5}{\nano\meter} and \SI{221.5}{\nano\meter} are attributed to the transitions from the $5p^6$ ground state to the $6p\left[\frac{5}{2}\right]_2$, $6p\left[\frac{3}{2}\right]_2$, $6p\left[\frac{1}{2}\right]_0$, $6p'\left[\frac{3}{2}\right]_2$ and $6p'\left[\frac{1}{2}\right]_0$ states, with the expected positions of the unperturbed transitions shown as vertical lines. For increasing buffer gas pressures, the respective peaks become increasingly broadened and shifted with respect to the unperturbed resonance position. Moreover for the lower and moderate helium pressure data a spectrally sharp peak, the line core, is visible which is shifted by a relatively small amount (approx. \SI{0.1}{\nano\meter}), on top of a broader spectral wing, with the latter becoming dominant for the \SI{80}{\bar} helium pressure data. The wings of the lines are asymmetrically broadened to the low wavelength side, which is evidence for a repulsive nature of the corresponding upper states xenon-helium potential curves.\\
	The bottom panel of Fig.~\ref{fig: excitationSpectrum} shows corresponding experimental data recorded with the use of krypton as a buffer gas. Noticeably, in this case observed resonances that can be clearly identified to allowed two-photon transitions are visible only for the lower and moderate buffer gas pressure regime (\SI{5}{\bar} and \SI{30}{\bar} data) for the $5p^6\rightarrow 5p^56p$ configuration. The corresponding lineshapes are here broadened and shifted towards higher wavelengths, as can be attributed to a attractive upper state xenon-krypton quasimolecular potential, which is in good agreement to observations in earlier work for the corresponding gas mixture at lower buffer gas pressure \cite{Freeman1977}. Remarkably, for a krypton buffer gas pressure of \SI{80}{\bar} the $5p^6\rightarrow 5p^56p$ resonances tend to vanish. Furthermore, with increasing pressure a weak resonance near \SI{245}{\nano\meter} wavelength appears, see also the inset, which can be attributed to dipole-forbidden transitions to the $5d\left[\frac{5}{2}\right]_3$ state. Corresponding observations have been reported in earlier works for the case of pure xenon gas \cite{Gornik1980}. Additionally, near \SI{231}{\nano\meter} wavelength a new resonance appears for both the \SI{30}{\bar} and the \SI{80}{\bar} krypton buffer gas data, which cannot be clearly assigned to a known transition. A possible explanation for this (as well as an alternative possibility for the observed weak line near \SI{245}{\nano\meter}) would be a molecular transition, as has been discussed for spectral features of the better theoretically characterized xenon-xenon system \cite{GornikTwoPhoton1981}. At the short-wavelength limit of the experimental xenon-krypton data, the $5p^56p'$ resonances are not clearly spectrally resolved even for the \SI{5}{\bar} data, with for wavelengths smaller than \SI{228}{\nano\meter} rather a continuum arising. For the higher krypton pressures, a further resonance seems to appear at the low wavelength-edge of our available excitation wavelength tuning range of \SI{220}{\nano\meter}. This could be attributed to another molecular transition appearing below \SI{220}{\nano\meter}.

\section{Probing the Kennard-Stepanov relation in a heteronuclear noble gas mixture} \label{sec: KennardStepanov}

	\begin{figure}
		\includegraphics[]{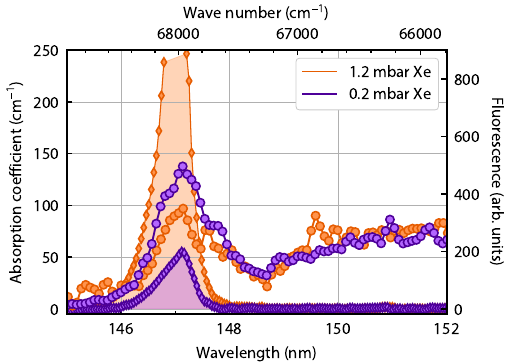}
		\caption{Absorption and fluorescence spectra for a xenon-krypton mixture for a krypton pressure of \SI{80}{\bar} and varying xenon pressures (see description in figure). The absorption data  are shown by (small) diamonds, and the area below the corresponding curves is indicated shaded. These spectra exhibit \color{black} a single peak centered  near \color{black} \SI{147}{\nano\meter} wavelength. For clarity, the absorption data has been capped above \SI{250}{\per\centi\meter} as the absorption in this range cannot be resolved due to an apparently vanishing transmission. The emission, with the corresponding data depicted by (large) circles, \color{black} extends to larger wavelengths.  In addition to the main peak, a second \color{black} characteristic feature  centered \color{black} around \SI{153}{\nano\meter}  resulting \color{black} from the emission of heteronuclear xenon-krypton excimers (similar as for the xenon-krypton emission spectra shown in Fig.~\ref{fig: rawSpectra}) that extends beyond the here-depicted wavelength range is observed. The in this plot used relatively narrow wavelength range was chosen to set a closer focus on the overlap region between absorption and emission.\color{black}\label{fig:absFluoOverlap}}
	\end{figure}

	\begin{figure}
		\includegraphics[]{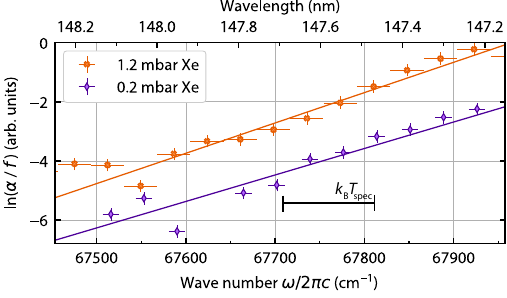}
		\caption{Logarithmic ratio of the absorption ($\alpha(\omega)$) and emission ($f(\omega)$) spectral profiles as a function of the optical wavenumber $\omega/2\pi c$, obtained by analyzing the experimental line shapes given in Fig.~\ref{fig:absFluoOverlap}  within the long-wavelength edge of the absorption line profile. \color{black} The lines are linear fits to the shown data in this wavelength range.\label{fig: kennardstepanov}}
	\end{figure}
	
	In this chapter, we present data examining the fulfillment of the Kennard-Stepanov relation for heteronuclear xenon-krypton gas mixtures. For this, both absorption and emission spectra of the $5p^6 \leftrightarrow 5p^56s$ one-photon transition of xenon subject to \SI{80}{\bar} of krypton buffer gas at different xenon concentrations have been recorded.\\
	In Fig.~\ref{fig:absFluoOverlap}, the connected purple diamonds (\SI{0.2}{\milli\bar} xenon partial pressure) and connected orange diamonds (\SI{1.2}{\milli\bar} xenon partial pressure)  with coloured filling underneath \color{black} show absorption spectra for corresponding gas mixtures. Further, the  connected circles without filling underneath give corresponding \color{black} experimental data for the spectrally resolved emission (following direct single-photon excitation, see Sec. \ref{sec: experimentalEnvironment}) for  the two gas mixtures \color{black} respectively. The resonances in  all \color{black} cases are around \SI{147}{\nano\meter} wavelength, with the emission showing a weakly Stokes-shifted main component that we associate to the atomic fluorescence, as well as a second, more red-shifted broader feature that is understood as stemming from the emission of xenon-krypton excimers (compare also with the corresponding discussion of the xenon-krypton data in Sec. \ref{sec: ExcitationSpectroscopy}) with the former component being more dominant for the lower xenon concentration.\\
	The shown emission data has been recorded following single-photon excitation using radiation near \SI{146}{\nano\meter}. Interestingly, the overall lineshape is relatively independent of the pathway of excitation; compare e.g.~the shape of the \SI{1.2}{\milli\bar} xenon partial pressure emission curve in Fig.~\ref{fig:absFluoOverlap} with corresponding data for the xenon-krypton mixture in Fig.~\ref{fig: rawSpectra}, as obtained by two-photon excitation. Comparable data as the emission data of Fig.~\ref{fig:absFluoOverlap} has also been obtained with different single-photon excitation wavelengths. This provides evidence for collisional redistribution in the quasimolecular potentials, being the essential prerequisite for the thermodynamic Kennard-Stepanov frequency scaling between absorption and emission to hold.\\
	To allow for a verification of the Kennard-Stepanov scaling from our (one-photon) spectroscopic data, \color{black} Fig.~\ref{fig: kennardstepanov} shows a plot of the logarithmic ratio of the absorption coefficient $\alpha(\omega)$ and the fluorescence signal $f(\omega)$ as a function of the optical wave number $\omega/2\pi c$. The data is plotted for a spectral range where the absorption and the emission lineshapes spectrally overlap on the longer wavelength edge of the absorption, see also Fig.~\ref{fig:absFluoOverlap}. The Kennard-Stepanov relation $\frac{f(\omega)}{\alpha(\omega)}\propto\omega^3\exp{\left(-\frac{\hbar\omega}{k_{\textrm{B}}T}\right)}$ in the used logarithmic representation predicts a linear dependence for the small wavelength range, with a slope that is determined by the inverse of the temperature.  Due to the logarithmic scaling, this temperature does not depend on an arbitrary scaling factor in the fluorescence signal. \color{black} The data points in Fig.~\ref{fig: kennardstepanov} are  the \color{black} corresponding experimental data. To this data a linear curve with a slope $hc/k_{\textrm{B}}T_\textrm{spec}$ has been fitted. The obtained spectral temperatures for xenon gas pressures of \SI{0.2}{\milli\bar} and \SI{1.2}{\milli\bar} are $T_\textrm{spec}=$ \SI{161}{\kelvin} and $T_\textrm{spec}=$ \SI{140}{\kelvin}, respectively. The obtained values for the spectral temperature are lower than room temperature, which is understood as a signature for incomplete thermalization of the quasimolecular manifolds. Note, however, the good agreement of the experimental data with a linear scaling over a range corresponding to $2k_{\textrm{B}}\cdot\SI{300}{\kelvin}$, or equivalently $4k_{\textrm{B}}\cdot T_\textrm{spec}$, where $T_\textrm{spec}$ is the spectral temperature. We thus expect that the noble-gas mixture is suitable for the thermalization of a vacuum-ultraviolet spectral regime photon gas in the described wavelength range.
	
\section{Conclusions}
\label{sec: conclusions}

	To conclude, we have reported experimental results on two-photon excitation spectroscopy of xenon-helium and xenon-krypton buffer gas mixtures in the high-pressure gaseous regime. For xenon-helium, $5p^6 \rightarrow 5p^56p$ and $5p^6 \rightarrow 5p^56p'$ transitions were observed for buffer gas pressures of up to \SI{80}{\bar}. In xenon-krypton, we observe stark differences between spectra at low and high pressure, where at high pressure not only a shift and broadening is apparent, but also new resonances appear. To achieve a more complete picture, in future measurements it would be interesting to extend the wavelength range of the excitation spectra to wavelengths smaller than our continuous tuning range of \SI{220}{\nano\meter} to in more detail investigate the behavior of the pressure broadened $5p^56p'$ resonances.\\
	For the case of xenon-krypton mixtures, we have also recorded both absorption and fluorescence spectra of the one-photon $5p^6 \rightarrow 5p^56s$ xenon resonance near \SI{147}{\nano\meter} wavelength at \SI{80}{\bar} buffer gas pressure, to test for the thermodynamic Kennard-Stepanov frequency scaling between absorption and emission. Our results show that such a frequency scaling is indeed fulfilled, though the observed spectral temperature is about a factor of two below the cell temperature. Overall, our experimental one- and two-photon spectroscopy data of the dense heteronuclear noble gas mixtures show stark differences depending on pressure and species of the buffer gas. This can provide valuable input for future calculations of molecular potentials.\\
	The experimental results furthermore show that dense heteronuclear noble gas mixtures are attractive candidate systems for a future photon Bose-Einstein condensate in the vacuum-ultraviolet spectral regime, where pumping could be achieved via two-photon excitation. While the spectral temperature observed at the maximum possible buffer gas pressure of \SI{80}{\bar} is below room temperature, future work should verify whether an increase of the collisional rate, as possible with an increased buffer gas pressure, leads to improved thermalization.  Additional insights could in future be obtained from time-resolved measurements \cite{Schmitt2015}, which on the one hand could provide a very direct further line of evidence for thermalization of the quasimolecular manifolds, as well as allow for a determination of the relevant timescales. \color{black} But we point out that already with the present  spectroscopic results on the Kennard-Stepanov relation\color{black},  see section \ref{sec: KennardStepanov}, \color{black} conditions for realization of a VUV photon Bose-Einstein condensate, which would exhibit a thermal cloud at the corresponding temperature, are favorable.
	
\begin{acknowledgments}
	We acknowledge support of the Deutsche Forschungsgemeinschaft within project WE 1748-25-1 (Grant No. 496090524) and major instrumentation grant INST217/916-1 FUGG (Grant No. 422472083) as well as SFB/TR 185 (Grant No. 277625399).
\end{acknowledgments}


\begin{thebibliography}{39}%
	\makeatletter
	\providecommand \@ifxundefined [1]{%
		\@ifx{#1\undefined}
	}%
	\providecommand \@ifnum [1]{%
		\ifnum #1\expandafter \@firstoftwo
		\else \expandafter \@secondoftwo
		\fi
	}%
	\providecommand \@ifx [1]{%
		\ifx #1\expandafter \@firstoftwo
		\else \expandafter \@secondoftwo
		\fi
	}%
	\providecommand \natexlab [1]{#1}%
	\providecommand \enquote  [1]{``#1''}%
	\providecommand \bibnamefont  [1]{#1}%
	\providecommand \bibfnamefont [1]{#1}%
	\providecommand \citenamefont [1]{#1}%
	\providecommand \href@noop [0]{\@secondoftwo}%
	\providecommand \href [0]{\begingroup \@sanitize@url \@href}%
	\providecommand \@href[1]{\@@startlink{#1}\@@href}%
	\providecommand \@@href[1]{\endgroup#1\@@endlink}%
	\providecommand \@sanitize@url [0]{\catcode `\\12\catcode `\$12\catcode
		`\&12\catcode `\#12\catcode `\^12\catcode `\_12\catcode `\%12\relax}%
	\providecommand \@@startlink[1]{}%
	\providecommand \@@endlink[0]{}%
	\providecommand \url  [0]{\begingroup\@sanitize@url \@url }%
	\providecommand \@url [1]{\endgroup\@href {#1}{\urlprefix }}%
	\providecommand \urlprefix  [0]{URL }%
	\providecommand \Eprint [0]{\href }%
	\providecommand \doibase [0]{https://doi.org/}%
	\providecommand \selectlanguage [0]{\@gobble}%
	\providecommand \bibinfo  [0]{\@secondoftwo}%
	\providecommand \bibfield  [0]{\@secondoftwo}%
	\providecommand \translation [1]{[#1]}%
	\providecommand \BibitemOpen [0]{}%
	\providecommand \bibitemStop [0]{}%
	\providecommand \bibitemNoStop [0]{.\EOS\space}%
	\providecommand \EOS [0]{\spacefactor3000\relax}%
	\providecommand \BibitemShut  [1]{\csname bibitem#1\endcsname}%
	\let\auto@bib@innerbib\@empty
	\bibitem [{\citenamefont {{Khalafinejad, S.}}\ \emph
		{et~al.}(2017)\citenamefont {{Khalafinejad, S.}}, \citenamefont {{von Essen,
				C.}}, \citenamefont {{Hoeijmakers, H. J.}}, \citenamefont {{Zhou, G.}},
		\citenamefont {{Klocová, T.}}, \citenamefont {{Schmitt, J. H. M. M.}},
		\citenamefont {{Dreizler, S.}}, \citenamefont {{Lopez-Morales, M.}},
		\citenamefont {{Husser, T.-O.}}, \citenamefont {{Schmidt, T. O. B.}},\ and\
		\citenamefont {{Collet, R.}}}]{Khalafinejad2017}%
	\BibitemOpen
	\bibfield  {author} {\bibinfo {author} {\bibnamefont {{Khalafinejad, S.}}},
		\bibinfo {author} {\bibnamefont {{von Essen, C.}}}, \bibinfo {author}
		{\bibnamefont {{Hoeijmakers, H. J.}}}, \bibinfo {author} {\bibnamefont
			{{Zhou, G.}}}, \bibinfo {author} {\bibnamefont {{Klocová, T.}}}, \bibinfo
		{author} {\bibnamefont {{Schmitt, J. H. M. M.}}}, \bibinfo {author}
		{\bibnamefont {{Dreizler, S.}}}, \bibinfo {author} {\bibnamefont
			{{Lopez-Morales, M.}}}, \bibinfo {author} {\bibnamefont {{Husser, T.-O.}}},
		\bibinfo {author} {\bibnamefont {{Schmidt, T. O. B.}}},\ and\ \bibinfo
		{author} {\bibnamefont {{Collet, R.}}},\ }\bibfield  {title} {\bibinfo
		{title} {Exoplanetary atmospheric sodium revealed by orbital motion -
			narrow-band transmission spectroscopy of {HD} 189733b with {UVES}},\ }\href
	{https://doi.org/10.1051/0004-6361/201629473} {\bibfield  {journal} {\bibinfo
			{journal} {A\&A}\ }\textbf {\bibinfo {volume} {598}},\ \bibinfo {pages}
		{A131} (\bibinfo {year} {2017})}\BibitemShut {NoStop}%
	\bibitem [{\citenamefont {{Allard, N. F.}}\ and\ \citenamefont {{Spiegelman,
				F.}}(2006)}]{Allard2006}%
	\BibitemOpen
	\bibfield  {author} {\bibinfo {author} {\bibnamefont {{Allard, N. F.}}}\ and\
		\bibinfo {author} {\bibnamefont {{Spiegelman, F.}}},\ }\bibfield  {title}
	{\bibinfo {title} {Collisional line profiles of rubidium and cesium perturbed
			by helium and molecular hydrogen},\ }\href
	{https://doi.org/10.1051/0004-6361:20054485} {\bibfield  {journal} {\bibinfo
			{journal} {A\&A}\ }\textbf {\bibinfo {volume} {452}},\ \bibinfo {pages} {351}
		(\bibinfo {year} {2006})}\BibitemShut {NoStop}%
	\bibitem [{\citenamefont {Leggett}\ \emph {et~al.}(2012)\citenamefont
		{Leggett}, \citenamefont {Saumon}, \citenamefont {Marley}, \citenamefont
		{Lodders}, \citenamefont {Canty}, \citenamefont {Lucas}, \citenamefont
		{Smart}, \citenamefont {Tinney}, \citenamefont {Homeier}, \citenamefont
		{Allard}, \citenamefont {Burningham}, \citenamefont {Day-Jones},
		\citenamefont {Fegley}, \citenamefont {Ishii}, \citenamefont {Jones},
		\citenamefont {Marocco}, \citenamefont {Pinfield},\ and\ \citenamefont
		{Tamura}}]{Leggett2012}%
	\BibitemOpen
	\bibfield  {author} {\bibinfo {author} {\bibfnamefont {S.~K.}\ \bibnamefont
			{Leggett}}, \bibinfo {author} {\bibfnamefont {D.}~\bibnamefont {Saumon}},
		\bibinfo {author} {\bibfnamefont {M.~S.}\ \bibnamefont {Marley}}, \bibinfo
		{author} {\bibfnamefont {K.}~\bibnamefont {Lodders}}, \bibinfo {author}
		{\bibfnamefont {J.}~\bibnamefont {Canty}}, \bibinfo {author} {\bibfnamefont
			{P.}~\bibnamefont {Lucas}}, \bibinfo {author} {\bibfnamefont {R.~L.}\
			\bibnamefont {Smart}}, \bibinfo {author} {\bibfnamefont {C.~G.}\ \bibnamefont
			{Tinney}}, \bibinfo {author} {\bibfnamefont {D.}~\bibnamefont {Homeier}},
		\bibinfo {author} {\bibfnamefont {F.}~\bibnamefont {Allard}}, \bibinfo
		{author} {\bibfnamefont {B.}~\bibnamefont {Burningham}}, \bibinfo {author}
		{\bibfnamefont {A.}~\bibnamefont {Day-Jones}}, \bibinfo {author}
		{\bibfnamefont {B.}~\bibnamefont {Fegley}}, \bibinfo {author} {\bibfnamefont
			{M.}~\bibnamefont {Ishii}}, \bibinfo {author} {\bibfnamefont {H.~R.~A.}\
			\bibnamefont {Jones}}, \bibinfo {author} {\bibfnamefont {F.}~\bibnamefont
			{Marocco}}, \bibinfo {author} {\bibfnamefont {D.~J.}\ \bibnamefont
			{Pinfield}},\ and\ \bibinfo {author} {\bibfnamefont {M.}~\bibnamefont
			{Tamura}},\ }\bibfield  {title} {\bibinfo {title} {The properties of the
			500{K} dwarf {UGPS} {J}072227.51 054031.2 and a study of the far-red flux of
			cold brown dwarfs},\ }\href {https://doi.org/10.1088/0004-637X/748/2/74}
	{\bibfield  {journal} {\bibinfo  {journal} {The Astrophysical Journal}\
		}\textbf {\bibinfo {volume} {748}},\ \bibinfo {pages} {74} (\bibinfo {year}
		{2012})}\BibitemShut {NoStop}%
	\bibitem [{\citenamefont {Uzdensky}\ and\ \citenamefont
		{Rightley}(2014)}]{Uzdensky2014}%
	\BibitemOpen
	\bibfield  {author} {\bibinfo {author} {\bibfnamefont {D.~A.}\ \bibnamefont
			{Uzdensky}}\ and\ \bibinfo {author} {\bibfnamefont {S.}~\bibnamefont
			{Rightley}},\ }\bibfield  {title} {\bibinfo {title} {Plasma physics of
			extreme astrophysical environments},\ }\href
	{https://doi.org/10.1088/0034-4885/77/3/036902} {\bibfield  {journal}
		{\bibinfo  {journal} {Reports on Progress in Physics}\ }\textbf {\bibinfo
			{volume} {77}},\ \bibinfo {pages} {036902} (\bibinfo {year}
		{2014})}\BibitemShut {NoStop}%
	\bibitem [{\citenamefont {Martin}(1960)}]{Martin1960}%
	\BibitemOpen
	\bibfield  {author} {\bibinfo {author} {\bibfnamefont {E.~A.}\ \bibnamefont
			{Martin}},\ }\bibfield  {title} {\bibinfo {title} {Experimental investigation
			of a high‐energy density, high‐pressure arc plasma},\ }\href
	{https://doi.org/10.1063/1.1735555} {\bibfield  {journal} {\bibinfo
			{journal} {Journal of Applied Physics}\ }\textbf {\bibinfo {volume} {31}},\
		\bibinfo {pages} {255} (\bibinfo {year} {1960})}\BibitemShut {NoStop}%
	\bibitem [{\citenamefont {Kennard}(1918)}]{KennardFluorescence1918}%
	\BibitemOpen
	\bibfield  {author} {\bibinfo {author} {\bibfnamefont {E.~H.}\ \bibnamefont
			{Kennard}},\ }\bibfield  {title} {\bibinfo {title} {On the thermodynamics of
			fluorescence},\ }\href {https://doi.org/10.1103/PhysRev.11.29} {\bibfield
		{journal} {\bibinfo  {journal} {Phys. Rev.}\ }\textbf {\bibinfo {volume}
			{11}},\ \bibinfo {pages} {29} (\bibinfo {year} {1918})}\BibitemShut {NoStop}%
	\bibitem [{\citenamefont {Stepanov}(1957)}]{StepanovFluorescence1957}%
	\BibitemOpen
	\bibfield  {author} {\bibinfo {author} {\bibfnamefont {B.}~\bibnamefont
			{Stepanov}},\ }\bibfield  {title} {\bibinfo {title} {Universal relation
			between the absorption spectra and luminescence spectra of complex
			molecules},\ }\href
	{https://www.mathnet.ru/php/archive.phtml?wshow=paper&jrnid=dan&paperid=21621&option_lang=eng}
	{\bibfield  {journal} {\bibinfo  {journal} {\textup{Dokl. Akad. Nauk SSSR.}}\
		}\textbf {\bibinfo {volume} {112}},\ \bibinfo {pages} {839} (\bibinfo {year}
		{1957})}\BibitemShut {NoStop}%
	\bibitem [{\citenamefont {Kasha}(1950)}]{KashaCharacterization1950}%
	\BibitemOpen
	\bibfield  {author} {\bibinfo {author} {\bibfnamefont {M.}~\bibnamefont
			{Kasha}},\ }\bibfield  {title} {\bibinfo {title} {Characterization of
			electronic transitions in complex molecules},\ }\href
	{https://doi.org/10.1039/DF9500900014} {\bibfield  {journal} {\bibinfo
			{journal} {Discuss. Faraday Soc.}\ }\textbf {\bibinfo {volume} {9}},\
		\bibinfo {pages} {14} (\bibinfo {year} {1950})}\BibitemShut {NoStop}%
	\bibitem [{\citenamefont {Moroshkin}\ \emph {et~al.}(2014)\citenamefont
		{Moroshkin}, \citenamefont {Weller}, \citenamefont {Sa\ss{}}, \citenamefont
		{Kl{\"a}rs},\ and\ \citenamefont {Weitz}}]{MoroshkinKennardStepanov2014}%
	\BibitemOpen
	\bibfield  {author} {\bibinfo {author} {\bibfnamefont {P.}~\bibnamefont
			{Moroshkin}}, \bibinfo {author} {\bibfnamefont {L.}~\bibnamefont {Weller}},
		\bibinfo {author} {\bibfnamefont {A.}~\bibnamefont {Sa\ss{}}}, \bibinfo
		{author} {\bibfnamefont {J.}~\bibnamefont {Kl{\"a}rs}},\ and\ \bibinfo
		{author} {\bibfnamefont {M.}~\bibnamefont {Weitz}},\ }\bibfield  {title}
	{\bibinfo {title} {{Kennard}-{Stepanov} {R}elation {C}onnecting {A}bsorption
			and {E}mission {S}pectra in an {A}tomic {G}as},\ }\href
	{https://doi.org/10.1103/PhysRevLett.113.063002} {\bibfield  {journal}
		{\bibinfo  {journal} {Phys. Rev. Lett.}\ }\textbf {\bibinfo {volume} {113}},\
		\bibinfo {pages} {063002} (\bibinfo {year} {2014})}\BibitemShut {NoStop}%
	\bibitem [{\citenamefont {Sawicki}\ and\ \citenamefont
		{Knox}(1996)}]{Sawicki1996}%
	\BibitemOpen
	\bibfield  {author} {\bibinfo {author} {\bibfnamefont {D.~A.}\ \bibnamefont
			{Sawicki}}\ and\ \bibinfo {author} {\bibfnamefont {R.~S.}\ \bibnamefont
			{Knox}},\ }\bibfield  {title} {\bibinfo {title} {Universal relationship
			between optical emission and absorption of complex systems: An alternative
			approach},\ }\href {https://doi.org/10.1103/PhysRevA.54.4837} {\bibfield
		{journal} {\bibinfo  {journal} {Phys. Rev. A}\ }\textbf {\bibinfo {volume}
			{54}},\ \bibinfo {pages} {4837} (\bibinfo {year} {1996})}\BibitemShut
	{NoStop}%
	\bibitem [{\citenamefont {van Roosbroeck}\ and\ \citenamefont
		{Shockley}(1954)}]{RoosbroeckShockley1954}%
	\BibitemOpen
	\bibfield  {author} {\bibinfo {author} {\bibfnamefont {W.}~\bibnamefont {van
				Roosbroeck}}\ and\ \bibinfo {author} {\bibfnamefont {W.}~\bibnamefont
			{Shockley}},\ }\bibfield  {title} {\bibinfo {title} {Photon-radiative
			recombination of electrons and holes in germanium},\ }\href
	{https://doi.org/10.1103/PhysRev.94.1558} {\bibfield  {journal} {\bibinfo
			{journal} {Phys. Rev.}\ }\textbf {\bibinfo {volume} {94}},\ \bibinfo {pages}
		{1558} (\bibinfo {year} {1954})}\BibitemShut {NoStop}%
	\bibitem [{\citenamefont {Kl{\"a}rs}\ \emph
		{et~al.}(2010{\natexlab{a}})\citenamefont {Kl{\"a}rs}, \citenamefont
		{Schmitt}, \citenamefont {Vewinger},\ and\ \citenamefont
		{Weitz}}]{KlaersCondensation2010}%
	\BibitemOpen
	\bibfield  {author} {\bibinfo {author} {\bibfnamefont {J.}~\bibnamefont
			{Kl{\"a}rs}}, \bibinfo {author} {\bibfnamefont {J.}~\bibnamefont {Schmitt}},
		\bibinfo {author} {\bibfnamefont {F.}~\bibnamefont {Vewinger}},\ and\
		\bibinfo {author} {\bibfnamefont {M.}~\bibnamefont {Weitz}},\ }\bibfield
	{title} {\bibinfo {title} {{Bose}-{Einstein} condensation of photons in an
			optical microcavity},\ }\href
	{https://doi.org/https://doi.org/10.1038/nature09567} {\bibfield  {journal}
		{\bibinfo  {journal} {Nature}\ }\textbf {\bibinfo {volume} {468}},\ \bibinfo
		{pages} {545} (\bibinfo {year} {2010}{\natexlab{a}})}\BibitemShut {NoStop}%
	\bibitem [{\citenamefont {Kl{\"a}rs}\ \emph
		{et~al.}(2010{\natexlab{b}})\citenamefont {Kl{\"a}rs}, \citenamefont
		{Vewinger},\ and\ \citenamefont {Weitz}}]{KlaersThermalization2010}%
	\BibitemOpen
	\bibfield  {author} {\bibinfo {author} {\bibfnamefont {J.}~\bibnamefont
			{Kl{\"a}rs}}, \bibinfo {author} {\bibfnamefont {F.}~\bibnamefont
			{Vewinger}},\ and\ \bibinfo {author} {\bibfnamefont {M.}~\bibnamefont
			{Weitz}},\ }\bibfield  {title} {\bibinfo {title} {Thermalization of a
			two-dimensional photonic gas in a ``white wall'' photon box},\ }\href
	{https://doi.org/https://doi.org/10.1038/nphys1680} {\bibfield  {journal}
		{\bibinfo  {journal} {Nat. Phys.}\ }\textbf {\bibinfo {volume} {6}},\
		\bibinfo {pages} {512} (\bibinfo {year} {2010}{\natexlab{b}})}\BibitemShut
	{NoStop}%
	\bibitem [{\citenamefont {Marelic}\ and\ \citenamefont
		{Nyman}(2015)}]{MarelicCondensation2015}%
	\BibitemOpen
	\bibfield  {author} {\bibinfo {author} {\bibfnamefont {J.}~\bibnamefont
			{Marelic}}\ and\ \bibinfo {author} {\bibfnamefont {R.~A.}\ \bibnamefont
			{Nyman}},\ }\bibfield  {title} {\bibinfo {title} {Experimental evidence for
			inhomogeneous pumping and energy-dependent effects in photon
			{B}ose-{E}instein condensation},\ }\href
	{https://doi.org/10.1103/PhysRevA.91.033813} {\bibfield  {journal} {\bibinfo
			{journal} {Phys. Rev. A}\ }\textbf {\bibinfo {volume} {91}},\ \bibinfo
		{pages} {033813} (\bibinfo {year} {2015})}\BibitemShut {NoStop}%
	\bibitem [{\citenamefont {Greveling}\ \emph {et~al.}(2018)\citenamefont
		{Greveling}, \citenamefont {Perrier},\ and\ \citenamefont {van
			Oosten}}]{GrevelingDensity2018}%
	\BibitemOpen
	\bibfield  {author} {\bibinfo {author} {\bibfnamefont {S.}~\bibnamefont
			{Greveling}}, \bibinfo {author} {\bibfnamefont {K.~L.}\ \bibnamefont
			{Perrier}},\ and\ \bibinfo {author} {\bibfnamefont {D.}~\bibnamefont {van
				Oosten}},\ }\bibfield  {title} {\bibinfo {title} {Density distribution of a
			{Bose}-{Einstein} condensate of photons in a dye-filled microcavity},\ }\href
	{https://doi.org/10.1103/PhysRevA.98.013810} {\bibfield  {journal} {\bibinfo
			{journal} {Phys. Rev. A}\ }\textbf {\bibinfo {volume} {98}},\ \bibinfo
		{pages} {013810} (\bibinfo {year} {2018})}\BibitemShut {NoStop}%
	\bibitem [{\citenamefont {Schmitt}\ \emph {et~al.}(2016)\citenamefont
		{Schmitt}, \citenamefont {Damm}, \citenamefont {Dung}, \citenamefont {Wahl},
		\citenamefont {Vewinger}, \citenamefont {Kl{\"a}rs},\ and\ \citenamefont
		{Weitz}}]{SchmittCoherence2016}%
	\BibitemOpen
	\bibfield  {author} {\bibinfo {author} {\bibfnamefont {J.}~\bibnamefont
			{Schmitt}}, \bibinfo {author} {\bibfnamefont {T.}~\bibnamefont {Damm}},
		\bibinfo {author} {\bibfnamefont {D.}~\bibnamefont {Dung}}, \bibinfo {author}
		{\bibfnamefont {C.}~\bibnamefont {Wahl}}, \bibinfo {author} {\bibfnamefont
			{F.}~\bibnamefont {Vewinger}}, \bibinfo {author} {\bibfnamefont
			{J.}~\bibnamefont {Kl{\"a}rs}},\ and\ \bibinfo {author} {\bibfnamefont
			{M.}~\bibnamefont {Weitz}},\ }\bibfield  {title} {\bibinfo {title}
		{{S}pontaneous {S}ymmetry {B}reaking and {P}hase {C}oherence of a {P}hoton
			{B}ose-{E}instein {C}ondensate {C}oupled to a {R}eservoir},\ }\href
	{https://doi.org/10.1103/PhysRevLett.116.033604} {\bibfield  {journal}
		{\bibinfo  {journal} {Phys. Rev. Lett.}\ }\textbf {\bibinfo {volume} {116}},\
		\bibinfo {pages} {033604} (\bibinfo {year} {2016})}\BibitemShut {NoStop}%
	\bibitem [{\citenamefont {Gerasimov}(2004)}]{Gerasimov2004}%
	\BibitemOpen
	\bibfield  {author} {\bibinfo {author} {\bibfnamefont {G.~N.}\ \bibnamefont
			{Gerasimov}},\ }\bibfield  {title} {\bibinfo {title} {Optical spectra of
			binary rare-gas mixtures},\ }\href
	{https://doi.org/10.1070/PU2004v047n02ABEH001681} {\bibfield  {journal}
		{\bibinfo  {journal} {Physics-Uspekhi}\ }\textbf {\bibinfo {volume} {47}},\
		\bibinfo {pages} {149} (\bibinfo {year} {2004})}\BibitemShut {NoStop}%
	\bibitem [{\citenamefont {Sansonetti}\ and\ \citenamefont
		{Martin}(2023)}]{NISTenergyLevels2023}%
	\BibitemOpen
	\bibfield  {author} {\bibinfo {author} {\bibfnamefont {J.}~\bibnamefont
			{Sansonetti}}\ and\ \bibinfo {author} {\bibfnamefont {W.}~\bibnamefont
			{Martin}},\ }\bibinfo {title} {Handbook of basic atomic spectroscopic data},\
	in\ \href {https://doi.org/https://dx.doi.org/10.18434/T4FW23} {\emph
		{\bibinfo {booktitle} {NIST Chemistry WebBook, Standard Reference Database
				No. 108}}}\ (\bibinfo  {publisher} {National Institute of Standards and
		Technology},\ \bibinfo {address} {Quantum Measurement Division, PML,
		Gaithersburg MD},\ \bibinfo {year} {2023})\BibitemShut {NoStop}%
	\bibitem [{\citenamefont {Zaidel}\ and\ \citenamefont
		{Shreider}(1970)}]{ZaidelVUV1970}%
	\BibitemOpen
	\bibfield  {author} {\bibinfo {author} {\bibfnamefont {A.}~\bibnamefont
			{Zaidel}}\ and\ \bibinfo {author} {\bibfnamefont {E.}~\bibnamefont
			{Shreider}},\ }\href@noop {} {\emph {\bibinfo {title} {Vacuum Ultraviolet
				Spectroscopy}}}\ (\bibinfo  {publisher} {Humphrey Science},\ \bibinfo
	{address} {Ann Arbor, London},\ \bibinfo {year} {1970})\BibitemShut {NoStop}%
	\bibitem [{\citenamefont {vom H\"ovel}\ \emph {et~al.}(2023)\citenamefont {vom
			H\"ovel}, \citenamefont {Huybrechts}, \citenamefont {Boltersdorf},
		\citenamefont {Wahl}, \citenamefont {Vewinger},\ and\ \citenamefont
		{Weitz}}]{Hoevel2023}%
	\BibitemOpen
	\bibfield  {author} {\bibinfo {author} {\bibfnamefont {T.}~\bibnamefont {vom
				H\"ovel}}, \bibinfo {author} {\bibfnamefont {F.}~\bibnamefont {Huybrechts}},
		\bibinfo {author} {\bibfnamefont {E.}~\bibnamefont {Boltersdorf}}, \bibinfo
		{author} {\bibfnamefont {C.}~\bibnamefont {Wahl}}, \bibinfo {author}
		{\bibfnamefont {F.}~\bibnamefont {Vewinger}},\ and\ \bibinfo {author}
		{\bibfnamefont {M.}~\bibnamefont {Weitz}},\ }\bibfield  {title} {\bibinfo
		{title} {Two-photon excitation and absorption spectroscopy of gaseous and
			supercritical xenon},\ }\href {https://doi.org/10.1103/PhysRevA.108.012821}
	{\bibfield  {journal} {\bibinfo  {journal} {Phys. Rev. A}\ }\textbf {\bibinfo
			{volume} {108}},\ \bibinfo {pages} {012821} (\bibinfo {year}
		{2023})}\BibitemShut {NoStop}%
	\bibitem [{\citenamefont {Ockenfels}\ \emph {et~al.}(2022)\citenamefont
		{Ockenfels}, \citenamefont {Roje}, \citenamefont {vom H{\"o}vel},
		\citenamefont {Vewinger},\ and\ \citenamefont
		{Weitz}}]{OckenfelsSpectroscopy2022}%
	\BibitemOpen
	\bibfield  {author} {\bibinfo {author} {\bibfnamefont {T.}~\bibnamefont
			{Ockenfels}}, \bibinfo {author} {\bibfnamefont {P.}~\bibnamefont {Roje}},
		\bibinfo {author} {\bibfnamefont {T.}~\bibnamefont {vom H{\"o}vel}}, \bibinfo
		{author} {\bibfnamefont {F.}~\bibnamefont {Vewinger}},\ and\ \bibinfo
		{author} {\bibfnamefont {M.}~\bibnamefont {Weitz}},\ }\bibfield  {title}
	{\bibinfo {title} {Spectroscopy of high-pressure rubidium-noble-gas
			mixtures},\ }\href {https://doi.org/10.1103/PhysRevA.106.012815} {\bibfield
		{journal} {\bibinfo  {journal} {Phys. Rev. A}\ }\textbf {\bibinfo {volume}
			{106}},\ \bibinfo {pages} {012815} (\bibinfo {year} {2022})}\BibitemShut
	{NoStop}%
	\bibitem [{\citenamefont {Wahl}\ \emph {et~al.}(2016)\citenamefont {Wahl},
		\citenamefont {Brausemann}, \citenamefont {Schmitt}, \citenamefont
		{Vewinger}, \citenamefont {Christopoulos},\ and\ \citenamefont
		{Weitz}}]{WahlAbsorption2018}%
	\BibitemOpen
	\bibfield  {author} {\bibinfo {author} {\bibfnamefont {C.}~\bibnamefont
			{Wahl}}, \bibinfo {author} {\bibfnamefont {R.}~\bibnamefont {Brausemann}},
		\bibinfo {author} {\bibfnamefont {J.}~\bibnamefont {Schmitt}}, \bibinfo
		{author} {\bibfnamefont {F.}~\bibnamefont {Vewinger}}, \bibinfo {author}
		{\bibfnamefont {S.}~\bibnamefont {Christopoulos}},\ and\ \bibinfo {author}
		{\bibfnamefont {M.}~\bibnamefont {Weitz}},\ }\bibfield  {title} {\bibinfo
		{title} {Absorption spectroscopy of xenon and ethylene--noble gas mixtures at
			high pressure: towards {B}ose--{E}instein condensation of vacuum ultraviolet
			photons},\ }\href
	{https://link.springer.com/article/10.1007/s00340-016-6566-x} {\bibfield
		{journal} {\bibinfo  {journal} {Appl. Phys. B}\ }\textbf {\bibinfo {volume}
			{122}},\ \bibinfo {pages} {296} (\bibinfo {year} {2016})}\BibitemShut
	{NoStop}%
	\bibitem [{\citenamefont {Wahl}\ \emph {et~al.}(2021)\citenamefont {Wahl},
		\citenamefont {Hoffmann}, \citenamefont {vom H{\"o}vel}, \citenamefont
		{Vewinger},\ and\ \citenamefont {Weitz}}]{WahlXenon2021}%
	\BibitemOpen
	\bibfield  {author} {\bibinfo {author} {\bibfnamefont {C.}~\bibnamefont
			{Wahl}}, \bibinfo {author} {\bibfnamefont {M.}~\bibnamefont {Hoffmann}},
		\bibinfo {author} {\bibfnamefont {T.}~\bibnamefont {vom H{\"o}vel}}, \bibinfo
		{author} {\bibfnamefont {F.}~\bibnamefont {Vewinger}},\ and\ \bibinfo
		{author} {\bibfnamefont {M.}~\bibnamefont {Weitz}},\ }\bibfield  {title}
	{\bibinfo {title} {Vacuum-ultraviolet absorption and emission spectroscopy of
			gaseous, liquid, and supercritical xenon},\ }\href
	{https://doi.org/10.1103/PhysRevA.103.022831} {\bibfield  {journal} {\bibinfo
			{journal} {Phys. Rev. A}\ }\textbf {\bibinfo {volume} {103}},\ \bibinfo
		{pages} {022831} (\bibinfo {year} {2021})}\BibitemShut {NoStop}%
	\bibitem [{\citenamefont {Borovich}\ \emph {et~al.}(1973)\citenamefont
		{Borovich}, \citenamefont {Zuev},\ and\ \citenamefont
		{Stavrovsky}}]{BorovichAbsorption1973}%
	\BibitemOpen
	\bibfield  {author} {\bibinfo {author} {\bibfnamefont {B.}~\bibnamefont
			{Borovich}}, \bibinfo {author} {\bibfnamefont {V.}~\bibnamefont {Zuev}},\
		and\ \bibinfo {author} {\bibfnamefont {D.}~\bibnamefont {Stavrovsky}},\
	}\bibfield  {title} {\bibinfo {title} {Pressure-induced ultraviolet
			absorption in rare gases: {Absorption} coefficients for mixtures of {Xe} and
			{Ar} at pressures up to 40 atm in the vicinity of 147 nm},\ }\href
	{https://doi.org/https://doi.org/10.1016/0022-4073(73)90037-X} {\bibfield
		{journal} {\bibinfo  {journal} {J. Quant. Spectrosc. Radiat. Transfer}\
		}\textbf {\bibinfo {volume} {13}},\ \bibinfo {pages} {1241} (\bibinfo {year}
		{1973})}\BibitemShut {NoStop}%
	\bibitem [{\citenamefont {Gornik}\ \emph {et~al.}(1981)\citenamefont {Gornik},
		\citenamefont {Kindt}, \citenamefont {Matthias},\ and\ \citenamefont
		{Schmidt}}]{GornikTwoPhoton1981}%
	\BibitemOpen
	\bibfield  {author} {\bibinfo {author} {\bibfnamefont {W.}~\bibnamefont
			{Gornik}}, \bibinfo {author} {\bibfnamefont {S.}~\bibnamefont {Kindt}},
		\bibinfo {author} {\bibfnamefont {E.}~\bibnamefont {Matthias}},\ and\
		\bibinfo {author} {\bibfnamefont {D.}~\bibnamefont {Schmidt}},\ }\bibfield
	{title} {\bibinfo {title} {Two-photon excitation of xenon atoms and dimers in
			the energy region of the $5{p}^{5}6p$ configuration},\ }\href
	{https://doi.org/10.1063/1.441856} {\bibfield  {journal} {\bibinfo  {journal}
			{J. Chem. Phys.}\ }\textbf {\bibinfo {volume} {75}},\ \bibinfo {pages} {68}
		(\bibinfo {year} {1981})}\BibitemShut {NoStop}%
	\bibitem [{\citenamefont {Raymond}\ \emph {et~al.}(1984)\citenamefont
		{Raymond}, \citenamefont {B{\"o}wering}, \citenamefont {Kuo},\ and\
		\citenamefont {Keto}}]{RaymondTwoPhoton1984}%
	\BibitemOpen
	\bibfield  {author} {\bibinfo {author} {\bibfnamefont {T.~D.}\ \bibnamefont
			{Raymond}}, \bibinfo {author} {\bibfnamefont {N.}~\bibnamefont
			{B{\"o}wering}}, \bibinfo {author} {\bibfnamefont {C.-Y.}\ \bibnamefont
			{Kuo}},\ and\ \bibinfo {author} {\bibfnamefont {J.~W.}\ \bibnamefont
			{Keto}},\ }\bibfield  {title} {\bibinfo {title} {Two-photon laser
			spectroscopy of xenon collision pairs},\ }\href
	{https://doi.org/10.1103/PhysRevA.29.721} {\bibfield  {journal} {\bibinfo
			{journal} {Phys. Rev. A}\ }\textbf {\bibinfo {volume} {29}},\ \bibinfo
		{pages} {721} (\bibinfo {year} {1984})}\BibitemShut {NoStop}%
	\bibitem [{\citenamefont {B{\"o}wering}\ \emph
		{et~al.}(1986{\natexlab{a}})\citenamefont {B{\"o}wering}, \citenamefont
		{Bruce},\ and\ \citenamefont {Keto}}]{BoweringCollisionalLifetimes1986}%
	\BibitemOpen
	\bibfield  {author} {\bibinfo {author} {\bibfnamefont {N.}~\bibnamefont
			{B{\"o}wering}}, \bibinfo {author} {\bibfnamefont {M.}~\bibnamefont
			{Bruce}},\ and\ \bibinfo {author} {\bibfnamefont {J.}~\bibnamefont {Keto}},\
	}\bibfield  {title} {\bibinfo {title} {Collisional deactivation of two-photon
			laser excited xenon $5{p}^{5}6p$. {II}. {Lifetimes} and total quench rates},\
	}\href {https://doi.org/10.1063/1.450568} {\bibfield  {journal} {\bibinfo
			{journal} {J. Chem. Phys.}\ }\textbf {\bibinfo {volume} {84}},\ \bibinfo
		{pages} {715} (\bibinfo {year} {1986}{\natexlab{a}})}\BibitemShut {NoStop}%
	\bibitem [{\citenamefont {B{\"o}wering}\ \emph
		{et~al.}(1986{\natexlab{b}})\citenamefont {B{\"o}wering}, \citenamefont
		{Bruce},\ and\ \citenamefont {Keto}}]{BoweringCollisionalState1986}%
	\BibitemOpen
	\bibfield  {author} {\bibinfo {author} {\bibfnamefont {N.}~\bibnamefont
			{B{\"o}wering}}, \bibinfo {author} {\bibfnamefont {M.}~\bibnamefont
			{Bruce}},\ and\ \bibinfo {author} {\bibfnamefont {J.}~\bibnamefont {Keto}},\
	}\bibfield  {title} {\bibinfo {title} {Collisional deactivation of two-photon
			laser excited xenon $5{p}^{5}6p$. {I}. {State}-to-state reaction rates},\
	}\href {https://doi.org/10.1063/1.450567} {\bibfield  {journal} {\bibinfo
			{journal} {J. Chem. Phys.}\ }\textbf {\bibinfo {volume} {84}},\ \bibinfo
		{pages} {709} (\bibinfo {year} {1986}{\natexlab{b}})}\BibitemShut {NoStop}%
	\bibitem [{\citenamefont {Whitehead}\ \emph {et~al.}(1995)\citenamefont
		{Whitehead}, \citenamefont {Pournasr}, \citenamefont {Bruce}, \citenamefont
		{Cai}, \citenamefont {Kohel}, \citenamefont {Layne},\ and\ \citenamefont
		{Keto}}]{WhiteheadDeactivation1995}%
	\BibitemOpen
	\bibfield  {author} {\bibinfo {author} {\bibfnamefont {C.~A.}\ \bibnamefont
			{Whitehead}}, \bibinfo {author} {\bibfnamefont {H.}~\bibnamefont {Pournasr}},
		\bibinfo {author} {\bibfnamefont {M.~R.}\ \bibnamefont {Bruce}}, \bibinfo
		{author} {\bibfnamefont {H.}~\bibnamefont {Cai}}, \bibinfo {author}
		{\bibfnamefont {J.}~\bibnamefont {Kohel}}, \bibinfo {author} {\bibfnamefont
			{W.~B.}\ \bibnamefont {Layne}},\ and\ \bibinfo {author} {\bibfnamefont
			{J.~W.}\ \bibnamefont {Keto}},\ }\bibfield  {title} {\bibinfo {title}
		{Deactivation of two-photon excited {Xe}($5p^5$$6p$,$6p^{\prime}$,$7p$) and
			{Kr}($4p^5$$5p$) in xenon and krypton},\ }\href
	{https://doi.org/10.1063/1.468763} {\bibfield  {journal} {\bibinfo  {journal}
			{J. Chem. Phys.}\ }\textbf {\bibinfo {volume} {102}},\ \bibinfo {pages}
		{1965} (\bibinfo {year} {1995})}\BibitemShut {NoStop}%
	\bibitem [{\citenamefont {Nowak}\ and\ \citenamefont
		{Fricke}(1985)}]{NowakHeteronuclear1985}%
	\BibitemOpen
	\bibfield  {author} {\bibinfo {author} {\bibfnamefont {G.}~\bibnamefont
			{Nowak}}\ and\ \bibinfo {author} {\bibfnamefont {J.}~\bibnamefont {Fricke}},\
	}\bibfield  {title} {\bibinfo {title} {The heteronuclear excimers {ArKr*},
			{ArXe*} and {KrXe*}},\ }\href {https://doi.org/10.1088/0022-3700/18/7/016}
	{\bibfield  {journal} {\bibinfo  {journal} {J. Phys. B}\ }\textbf {\bibinfo
			{volume} {18}},\ \bibinfo {pages} {1355} (\bibinfo {year}
		{1985})}\BibitemShut {NoStop}%
	\bibitem [{\citenamefont {Gerasimov}\ \emph {et~al.}(2000)\citenamefont
		{Gerasimov}, \citenamefont {Volkova}, \citenamefont {Hallin}, \citenamefont
		{Zvereva},\ and\ \citenamefont {Heikensheld}}]{Gerasimov2000}%
	\BibitemOpen
	\bibfield  {author} {\bibinfo {author} {\bibfnamefont {G.}~\bibnamefont
			{Gerasimov}}, \bibinfo {author} {\bibfnamefont {G.}~\bibnamefont {Volkova}},
		\bibinfo {author} {\bibfnamefont {R.}~\bibnamefont {Hallin}}, \bibinfo
		{author} {\bibfnamefont {G.}~\bibnamefont {Zvereva}},\ and\ \bibinfo {author}
		{\bibfnamefont {F.}~\bibnamefont {Heikensheld}},\ }\bibfield  {title}
	{\bibinfo {title} {Vuv spectrum of the barrier discharge in a krypton-xenon
			mixture},\ }\href {https://doi.org/https://doi.org/10.1134/1.626884}
	{\bibfield  {journal} {\bibinfo  {journal} {Atomic Spectroscopy}\ }\textbf
		{\bibinfo {volume} {88}},\ \bibinfo {pages} {814} (\bibinfo {year}
		{2000})}\BibitemShut {NoStop}%
	\bibitem [{\citenamefont {Efthimiopoulos}\ \emph {et~al.}(1997)\citenamefont
		{Efthimiopoulos}, \citenamefont {Zouridis},\ and\ \citenamefont
		{Ulrich}}]{EfthimiopoulosExcimer1997}%
	\BibitemOpen
	\bibfield  {author} {\bibinfo {author} {\bibfnamefont {T.}~\bibnamefont
			{Efthimiopoulos}}, \bibinfo {author} {\bibfnamefont {D.}~\bibnamefont
			{Zouridis}},\ and\ \bibinfo {author} {\bibfnamefont {A.}~\bibnamefont
			{Ulrich}},\ }\bibfield  {title} {\bibinfo {title} {Excimer emission spectra
			of rare gas mixtures using either a supersonic expansion or a heavy-ion-beam
			excitation},\ }\href {https://doi.org/10.1088/0022-3727/30/12/010} {\bibfield
		{journal} {\bibinfo  {journal} {J. Phys. D}\ }\textbf {\bibinfo {volume}
			{30}},\ \bibinfo {pages} {1746} (\bibinfo {year} {1997})}\BibitemShut
	{NoStop}%
	\bibitem [{\citenamefont {Freeman}\ \emph {et~al.}(1977)\citenamefont
		{Freeman}, \citenamefont {Yohino},\ and\ \citenamefont
		{Tanaka}}]{Freeman1977}%
	\BibitemOpen
	\bibfield  {author} {\bibinfo {author} {\bibfnamefont {D.~E.}\ \bibnamefont
			{Freeman}}, \bibinfo {author} {\bibfnamefont {K.}~\bibnamefont {Yohino}},\
		and\ \bibinfo {author} {\bibfnamefont {Y.}~\bibnamefont {Tanaka}},\
	}\bibfield  {title} {\bibinfo {title} {Vacuum ultraviolet absorption spectra
			of binary rare gas mixtures and the properties of heteronuclear rare gas van
			der waals molecules},\ }\href
	{https://doi.org/https://doi.org/10.1063/1.435343} {\bibfield  {journal}
		{\bibinfo  {journal} {J. Chem. Phys.}\ }\textbf {\bibinfo {volume} {67}},\
		\bibinfo {pages} {3462} (\bibinfo {year} {1977})}\BibitemShut {NoStop}%
	\bibitem [{\citenamefont {Gray}\ \emph {et~al.}(0217)\citenamefont {Gray},
		\citenamefont {Bosert}, \citenamefont {Shyur}, \citenamefont {Saarel},
		\citenamefont {Briles},\ and\ \citenamefont {Lewandowski}}]{Gray2021}%
	\BibitemOpen
	\bibfield  {author} {\bibinfo {author} {\bibfnamefont {J.~M.}\ \bibnamefont
			{Gray}}, \bibinfo {author} {\bibfnamefont {J.}~\bibnamefont {Bosert}},
		\bibinfo {author} {\bibfnamefont {Y.}~\bibnamefont {Shyur}}, \bibinfo
		{author} {\bibfnamefont {B.}~\bibnamefont {Saarel}}, \bibinfo {author}
		{\bibfnamefont {T.~C.}\ \bibnamefont {Briles}},\ and\ \bibinfo {author}
		{\bibfnamefont {H.}~\bibnamefont {Lewandowski}},\ }\bibfield  {title}
	{\bibinfo {title} {Characterization of a vacuum ultraviolet light source at
			118nm},\ }\href {https://doi.org/https://doi.org/10.1063/5.0033135}
	{\bibfield  {journal} {\bibinfo  {journal} {J. Chem. Phys.}\ }\textbf
		{\bibinfo {volume} {154}},\ \bibinfo {pages} {024201} (\bibinfo {year}
		{20217})}\BibitemShut {NoStop}%
	\bibitem [{\citenamefont {Bruce}\ \emph {et~al.}(1990)\citenamefont {Bruce},
		\citenamefont {Layne}, \citenamefont {Whitehead},\ and\ \citenamefont
		{Keto}}]{BruceRates1990}%
	\BibitemOpen
	\bibfield  {author} {\bibinfo {author} {\bibfnamefont {M.~R.}\ \bibnamefont
			{Bruce}}, \bibinfo {author} {\bibfnamefont {W.~B.}\ \bibnamefont {Layne}},
		\bibinfo {author} {\bibfnamefont {C.~A.}\ \bibnamefont {Whitehead}},\ and\
		\bibinfo {author} {\bibfnamefont {J.~W.}\ \bibnamefont {Keto}},\ }\bibfield
	{title} {\bibinfo {title} {Radiative lifetimes and collisional deactivation
			of two-photon excited xenon in argon and xenon},\ }\href
	{https://doi.org/10.1063/1.457939} {\bibfield  {journal} {\bibinfo  {journal}
			{J. Chem. Phys.}\ }\textbf {\bibinfo {volume} {92}},\ \bibinfo {pages} {2917}
		(\bibinfo {year} {1990})}\BibitemShut {NoStop}%
	\bibitem [{\citenamefont {Ledru}\ \emph {et~al.}(2007)\citenamefont {Ledru},
		\citenamefont {Marchal}, \citenamefont {Merbahi}, \citenamefont {Gardou},\
		and\ \citenamefont {Sewraj}}]{Ledru2007}%
	\BibitemOpen
	\bibfield  {author} {\bibinfo {author} {\bibfnamefont {G.}~\bibnamefont
			{Ledru}}, \bibinfo {author} {\bibfnamefont {F.}~\bibnamefont {Marchal}},
		\bibinfo {author} {\bibfnamefont {N.}~\bibnamefont {Merbahi}}, \bibinfo
		{author} {\bibfnamefont {J.~P.}\ \bibnamefont {Gardou}},\ and\ \bibinfo
		{author} {\bibfnamefont {N.}~\bibnamefont {Sewraj}},\ }\bibfield  {title}
	{\bibinfo {title} {Study of the formation and decay of {KrXe}* excimers at
			room temperature following selective excitation of the xenon 6s states},\
	}\href {https://doi.org/10.1088/0953-4075/40/10/002} {\bibfield  {journal}
		{\bibinfo  {journal} {Journal of Physics B: Atomic, Molecular and Optical
				Physics}\ }\textbf {\bibinfo {volume} {40}},\ \bibinfo {pages} {1651}
		(\bibinfo {year} {2007})}\BibitemShut {NoStop}%
	\bibitem [{\citenamefont
		{G{\"o}ppert-Mayer}(1931)}]{GoeppertMayerQuadratic1931}%
	\BibitemOpen
	\bibfield  {author} {\bibinfo {author} {\bibfnamefont {M.}~\bibnamefont
			{G{\"o}ppert-Mayer}},\ }\bibfield  {title} {\bibinfo {title} {{\"U}ber
			{E}lementarakte mit zwei {Q}uantenspr{\"u}ngen},\ }\href
	{https://doi.org/https://doi.org/10.1002/andp.19314010303} {\bibfield
		{journal} {\bibinfo  {journal} {Annal. Phys.}\ }\textbf {\bibinfo {volume}
			{401}},\ \bibinfo {pages} {273} (\bibinfo {year} {1931})}\BibitemShut
	{NoStop}%
	\bibitem [{\citenamefont {Gornik}\ \emph {et~al.}(1980)\citenamefont {Gornik},
		\citenamefont {Kindt}, \citenamefont {Matthias}, \citenamefont {Rinneberg},\
		and\ \citenamefont {Schmidt}}]{Gornik1980}%
	\BibitemOpen
	\bibfield  {author} {\bibinfo {author} {\bibfnamefont {W.}~\bibnamefont
			{Gornik}}, \bibinfo {author} {\bibfnamefont {S.}~\bibnamefont {Kindt}},
		\bibinfo {author} {\bibfnamefont {E.}~\bibnamefont {Matthias}}, \bibinfo
		{author} {\bibfnamefont {H.}~\bibnamefont {Rinneberg}},\ and\ \bibinfo
		{author} {\bibfnamefont {D.}~\bibnamefont {Schmidt}},\ }\bibfield  {title}
	{\bibinfo {title} {Off-resonant $e2$ transition observed in two-photon
			absorption in {Xe I}},\ }\href {https://doi.org/10.1103/PhysRevLett.45.1941}
	{\bibfield  {journal} {\bibinfo  {journal} {Phys. Rev. Lett.}\ }\textbf
		{\bibinfo {volume} {45}},\ \bibinfo {pages} {1941} (\bibinfo {year}
		{1980})}\BibitemShut {NoStop}%
	\bibitem [{\citenamefont {Schmitt}\ \emph {et~al.}(2015)\citenamefont
		{Schmitt}, \citenamefont {Damm}, \citenamefont {Dung}, \citenamefont
		{Vewinger}, \citenamefont {Klaers},\ and\ \citenamefont
		{Weitz}}]{Schmitt2015}%
	\BibitemOpen
	\bibfield  {author} {\bibinfo {author} {\bibfnamefont {J.}~\bibnamefont
			{Schmitt}}, \bibinfo {author} {\bibfnamefont {T.}~\bibnamefont {Damm}},
		\bibinfo {author} {\bibfnamefont {D.}~\bibnamefont {Dung}}, \bibinfo {author}
		{\bibfnamefont {F.}~\bibnamefont {Vewinger}}, \bibinfo {author}
		{\bibfnamefont {J.}~\bibnamefont {Klaers}},\ and\ \bibinfo {author}
		{\bibfnamefont {M.}~\bibnamefont {Weitz}},\ }\bibfield  {title} {\bibinfo
		{title} {Thermalization kinetics of light: From laser dynamics to equilibrium
			condensation of photons},\ }\href
	{https://doi.org/10.1103/PhysRevA.92.011602} {\bibfield  {journal} {\bibinfo
			{journal} {Phys. Rev. A}\ }\textbf {\bibinfo {volume} {92}},\ \bibinfo
		{pages} {011602} (\bibinfo {year} {2015})}\BibitemShut {NoStop}%
\end{thebibliography}
\end{document}